\def\rfr#1{eq.(\ref{#1})}

\def\leti{Lense--Thirring}

\def\eqi{\begin{equation}}
\def\eqf{\end{equation}}
\def\eqia{\begin{eqnarray}}
\def\eqfa{\end{eqnarray}}

\def\rp#1#2{{#1\over#2}}

\def\mlt{{\rm \mu_{LT}}}

\def\lb#1{\label{#1}}
\def\lg{LAGEOS}
\def\lgg{LAGEOS II}

\def\bm#1{{\mbox{\boldmath$#1$\unboldmath}}}

\documentclass[12pt]{iopart}
\begin{document}

\title[A LAGEOS--LAGEOS II--OPTIS mission]{On the possibility of measuring
the Lense--Thirring effect with a LAGEOS--LAGEOS
II--OPTIS--mission}

\author{Lorenzo Iorio\dag \footnote[1]{To whom correspondence should
be addressed (Lorenzo.Iorio@ba.infn.it)}, Ignazio Ciufolini\ddag,
Erricos C Pavlis\S, Stephan Schiller+, Hansj\"{o}rg Dittus\P\ and
Claus L\"{a}mmerzahl\P }

\address{\dag\ Dipartimento di Fisica dell'Universit${\rm \grave{a}}$ di Bari, Via Amendola 173, 70126, Bari,
Italy}

\address{\ddag\ Dipartimento di Ingegneria dell'Innovazione
dell'Universit{\`{a}} di Lecce and INFN Sezione di Lecce, via
Monteroni, 73100, Lecce, Italy}

\address{\S\ Joint Center for
Earth Systems Technology (JCET/UMBC), University of Maryland,
Baltimore County, 1000 Hilltop Circle, Baltimore, Maryland, USA
21250}

\address{+\ Institute for Experimental Physics,
Heinrich-Heine-University D\"{u}sseldorf, 40225 D\"{u}sseldorf,
Germany}

\address{\P\ ZARM, University of Bremen, 28359 Bremen,
Germany}

\begin{abstract}
A space mission, OPTIS, has been proposed for testing the
foundations of Special Relativity and Post--Newtonian gravitation
in the field of Earth. The constraints posed on the original OPTIS
orbital geometry would allow for a rather wide range of
possibilities for the final OPTIS orbital parameters. This freedom
could be exploited for further tests of Post-Newtonian gravity. In
this paper we wish to preliminarily investigate if it would be
possible to use the orbital data from OPTIS together with those
from the existing geodetic passive laser ranged LAGEOS and LAGEOS
II satellites in order to perform precise measurements of the
Lense--Thirring effect. With regard to this possibility, it is
important to notice that the drag--free technology which should be
adopted for the OPTIS mission would yield a lifetime of many years
for this satellite. It turns out that the best choice would
probably be to adopt the same orbital configuration of the
proposed LAGEOS--like LARES satellite and, for testing, select a
linear combination including the nodes of LAGEOS, LAGEOS II and
OPTIS and the perigee of OPTIS. The total systematic error should
be of the order of $1\%$. The LARES orbital geometry should not be
too in conflict with the original specifications of the OPTIS
mission. However, a compromise solution could be adopted as well.
A comparison with the new perspectives of measuring the
Lense--Thirring effect with the existing laser--tracked satellites
opened by the new gravity models from CHAMP and, especially, GRACE
is made. It turns out that an OPTIS/LARES mission would still be
of great significance because the obtainable accuracy would be
better than that offered by a reanalysis of the currently existing
satellites.
\end{abstract}

\section{Introduction}
Since gravity is the by far most weak interaction, tests of
relativistic gravity always go to the limits of experimental
capabilities. One way to increase the accuracy of results is to go
to space where much larger distances, velocities, gravitational
potential differences, and, most important, free fall for an, in
principle,  infinitely long time are available. These conditions
are the experimental basis for the gravity-- and
relativity--related space missions GP-A (Vessot $et\ al$ 1980),
GP-B (Everitt $et\ al$ 2001), GG (Nobili $et\ al$ 2000),
MICROSCOPE (Touboul 2001b), STEP (Lockerbie $et\ al$ 2001),
SPACETIME (Maleki and Prestage 2001), HYPER (see on the WEB ${\rm
http://www.esa.int/export/esaSC/SEM056WO4HD\_index\_0\_m.html}$),
ASTROD (Huang $et\ al$ 2002), LATOR (Turyshev $et\ al$ 2003a;
2003b), SEE (Sanders $et\ al$ 2000), and OPTIS (L${\rm
\ddot{a}}$mmerzahl $et\ al$ 2001) and the observations based on
the LAGEOS and LAGEOS II system (Ciufolini 2002). (See also
(L${\rm \ddot{a}}$mmerzahl and Dittus 2002) for a review).

In this paper we consider the mission OPTIS (L${\rm
\ddot{a}}$mmerzahl $et\ al$ 2001) which was designed for much
improved tests of (i) the independence of the velocity of light
from the velocity of the laboratory, and (ii) of the universality
of the gravitational redshift. We claim that this mission might
furthermore be used for an improvement of the tests of the
Post--Newtonian gravitomagnetic Lense--Thirring effect (Lense and
Thirring 1918) which until now has been experimentally checked
with an accuracy of the order of\footnote{Other scientists in
(Ries $et\ al$ 2003) propose a different error budget.} 20\%--30\%
(Ciufolini 2002) by analyzing the laser data to the existing
geodetic satellites LAGEOS and LAGEOS II. Therefore, there is a
tempting requirement to improve the quality of this test which
OPTIS may contribute to. The main reason why the OPTIS mission may
contribute to an improvement of the test of the Lense--Thirring
effect is the drag--free motion of the satellite which is made
possible due to very accurate inertial sensors and very
fine--tunable thrusters.


A test of gravitomagnetic effects is the subject of many space
missions: the LAGEOS--LAGEOS II system already observed the
Lense--Thirring effect in its original form, namely through the
precession of the satellite's node $\Omega$ and pericentre
$\omega$. GP-B
aims at a test of the Lense--Thirring effect in the form of the
frame--dagging of inertial systems represented by gyroscopes, also
called Schiff effect (Schiff 1960) (see (Sch${\rm\ddot{a}}$fer
2003) for a short review). This is a local version of
gravitomagnetism since in this case the phenomenon is not related
to a whole orbit but to a small region of space. In both cases the
effect is related to the $g_{0i}$ part of the spacetime metric.
This also extends to the usually applied PPN parameterization.
In this sense, also the HYPER mission is planned to test the local
frame dragging while OPTIS and ASTROD are sensitive to the global
effect. Another realization of a global version of gravitomagnetic
effects is based on clocks in counter--rotating satellites
(Mashhoon $et\ al$ 2001) which, however, is beyond today's
technical capabilities.

In this paper we first briefly review the basic features of the
OPTIS mission and the possible tests of relativistic gravity using
satellites. Then we consider the features of a combined
LAGEOS--LAGEOS II--OPTIS scheme in order to observe the
Lense--Thirring effect. The expected accuracy of the observation
of the Lense--Thirring effect for various scenarios for the OPTIS
mission are calculated and compared.
\section{OPTIS - a satellite mission for testing basic aspects of Special and General Relativity}
OPTIS  (L${\rm \ddot{a}}$mmerzahl $et\ al$ 2001) is a recently
proposed satellite--based mission\footnote{See also
http://www.exphy.uni-duesseldorf.de/OPTIS/optis.html.} which would
allow for precise tests of basic principles underlying Special and
General Relativity. This mission is based on the use of a spinning
drag--free satellite in an eccentric, high--altitude orbit which
should allow to perform a three orders of magnitude improved
Michelson--Morley test and a two orders of magnitude improved
Kennedy--Thorndike test. Moreover, it should also be possible to
improve by two orders of magnitude the tests of the universality
of the gravitational redshift by comparison of an atomic clock
with an optical clock. The proposed experiments are based on
ultrastable optical cavities, lasers, an atomic clock and a
frequency comb generator. Since it is not particularly important
for the present version of the mission the final orbital
configuration of OPTIS has not yet been fixed; in (L${\rm
\ddot{a}}$mmerzahl $et\ al$ 2001) a perigee height of 10000 km and
apogee height of 36000 km, with respect to Earth's surface, are
provisionally proposed assuming a launch with Ariane 5.

The requirements posed by the drag--free technology to be used,
based on the field emission electrical propulsion (FEEP) concept,
yield orbital altitudes not less than 1000 km. On the other hand,
the eccentricity should not be too high in order to prevent
passage in the Van Allen belts which could affect the on--board
capacitive reference sensor. Moreover, the orbital period $P_{\rm
OPT}$ should be shorter than the Earth's daily rotation of 24
hours. The orbital configuration proposed in (L${\rm
\ddot{a}}$mmerzahl $et\ al$ 2001) would imply a semimajor axis
$a_{\rm OPT}=29300$ km and an eccentricity $e_{\rm OPT}=0.478$.
With such values the difference of the gravitational potential
$U$, which is relevant for the gravitational redshift test, would
amount to \eqi \rp{\Delta U}{c^2}=\rp{GM}{c^2
a}\left[\rp{1}{(1-e)}-\rp{1}{(1+e)}\right]\sim 1.8\times
10^{-10},\lb{reds} \eqf where $G$ is the Newtonian gravitational
constant, $M$ the mass of Earth and $c$ the speed of light in
vacuum. The result of \rfr{reds} is about three orders of
magnitude better than that obtainable in an Earth--based
experiment.

An essential feature of OPTIS is the drag--free control of the
orbit. Drag--free motion is required for the SR and GR tests which
are carried through using optical resonators. Even very small
residual accelerations of $10^{-7}\, g$ may distort the resonators
leading to error signals. As a by--product, this drag--free
control also guarantees a very high quality geodesic motion which
may be used, when being tracked, as probe of orbital relativistic
gravitational effects.

For a drag--free motion of the satellite a sensor measuring the
actual acceleration and thrusters counteracting any acceleration
to the required precision are needed. The sensor, which is based
on a capacitive determination of the position of a test mass, has
a sensitivity of up to $10^{-12} \hbox{cm}\ {\hbox{s}}^{-2} {\rm
Hz}^{-\rp{1}{2}}$ (Touboul 2001a). This means that for one orbit
of about 12 h the difference of the real position from the
position achieved by ideal drag--free motion is of the order of 2
mm. Similar drag--free systems of similar accuracy and with
mission adapted modifications will be used in MICROSCOPE, STEP and
LISA. These systems have a lifetime of many years.
\section{Review of possible satellite tests of Post--Newtonian gravitation}
As stated above, the observation of the motion of freely falling
test bodies, i.e. satellites, is an important and feasible way to
test Post--Newtonian properties of the gravitational field created
by Earth. These are tests of the gravitational redshift and of
general relativistic gravitoelectromagnetic effects such as the
Lense--Thirring effect, all of order $\mathcal{O}(c^{-2})$.
\subsection{The gravitational redshift}
The gravitational redshift relates the frequencies $f$ of clocks
located at different gravitational potentials to the potential
difference
\begin{equation}
\frac{\Delta f}{f} = \frac{f(\bm r) - f({\bm r}_0)}{f({\bm r}_0)}
= \psi_{\rm clock}\frac{U(\bm r) - U(\bm r _0)}{c^2} \, .
\end{equation}
In General Relativity $\psi_{\rm clock} = 1$. This has been
tested, e.g., by Pound and Rebka (Pound and Rebka 1960) and at
best by the first fundamental physics space mission GP-A (Vessot
$et\ al$ 1980) with an accuracy $|\psi_{\rm clock} - 1| \leq 1.4
\times 10^{-4}$.

OPTIS aims at a test of this gravitational redshift with up to
three orders of improvement, that is up to $|\psi_{\rm clock} - 1|
\leq 10^{-7}$. In comparison, the planned experiments ACES-PHARAO,
SUMO, and PARCS to be carried through onboard of the International
Space Station (ISS) (L\"ammerzahl {\it et al} 2004) are supposed
to reach the $10^{-5}$ level, while the ISS project RACE is
projected to approach the $10^{-7}$ level. As compared to the ISS,
OPTIS has the advantage to fly on a high elliptic orbit, to have a
long mission time and to have verious clocks onboard (H--maser,
optical resonators, ion clocks). The OPTIS mission also aims at a
test of the universality of this gravitational redshift, that is
the equality of $\psi_{\rm clock}$ for different clocks,
$|\psi_{\rm clock2} - \psi_{\rm clock1}|$. Due to the various
clocks onboard of the OPTIS satellite also various combinations
can be tested. Again, the high elliptic orbit, the mission time as
well as the number of clocks will lead to an improvement of
previous results by up to three orders.
\subsection{The Lense--Thirring effect}
One of the most interesting Post--Newtonian gravitational effects
is the general relativistic gravitomagnetic Lense--Thirring effect
or dragging of inertial frames whose source is the proper angular
momentum \bm J of the central mass which acts as source of the
gravitational field. Its effect on the orientations of the spins
{\bm s} of four freely orbiting superconducting gyroscopes should
be tested, among other things, by the important GP--B mission
(Everitt $et\ al$ 2001) at a claimed accuracy level of the order
of 1$\%$ or better. Another possible way to measure such
relativistic effect is the analysis of the laser--ranged data of
some existing, or proposed, geodetic satellites of LAGEOS--type,
such as LAGEOS, LAGEOS II (Ciufolini 1996) and the proposed LAGEOS
III--LARES (Ciufolini 1986; 1998). In this case the whole orbit of
the satellite is to be thought of as a giant gyroscope whose node
and perigee undergo the Lense--Thirring secular precessions
\begin{eqnarray}
\dot\Omega_{\rm LT} &=&\rp{2GJ}{c^2 a^3(1-e^2)^{\rp{3}{2}}},\\
\dot\omega_{\rm LT} &=&-\rp{6GJ\cos i}{c^2
a^3(1-e^2)^{\rp{3}{2}}},
\end{eqnarray} where $i$ is the inclination of
the orbital plane to the Earth's equator. Note that in the
original paper by Lense and Thirring the longitude of the
pericentre $\varpi=\Omega+\omega$ is used instead of $\omega$.

Since 1996 measurements of the Lense-Thirring dragging of the
orbits of the existing LAGEOS and LAGEOS II satellites at a
claimed accuracy of the order of 20$\%$--30$\%$ (Ciufolini $et\
al$ 1998; Ciufolini 2002) have been reported.
Based on the original proposal of a laser ranging mission LARES
(LAser RElativity Satellite), recently an improved version of this
mission has been proposed  (Iorio $et\ al$ 2002; Iorio 2003a)
which should allow to reach an accuracy level of the order of
$1\%$. Below we are going to discuss possible measurement of the
Lense--Thirring effect by tracking the OPTIS satellite in the
LARES orbital configuration.
%
%
\section{A joint LAGEOS--LAGEOS II--OPTIS relativity measurement}
In this paper we wish to investigate the possibility to use the
orbital data of OPTIS for performing precise tests of general
relativistic gravitoelectromagnetism as well. The rather free
choice of the orbital parameters of OPTIS and the use of a new
drag--free technology open up the possibility to extend its
scientific significance with new important general relativistic
gravitoelectromagnetic tests.
Indeed, it would be of great impact and scientific significance to
concentrate as many relativistic tests as possible in a single
mission, including also measurements in geodesy, geodynamics.
Another important point is that OPTIS is currently under serious
examination by a national space agency-the German DLR. Then, even
if it turns out that OPTIS would yield little or no advantages for
the measurement of the Lense--Thirring effect with respect to the
originally proposed LARES, if it will be finally approved and
launched it will nevertheless be a great chance for detecting,
among other things, the Lense-Thirring effect.

In Table \ref{para} we report the orbital parameters of the
existing or proposed LAGEOS--type satellites and of the originally
proposed OPTIS configuration.
\begin{table}
\caption{Orbital parameters of \lg, \lgg, LARES and OPTIS.}
\label{para} \lineup
\begin{tabular}{@{}lllll}
\br
Orbital parameter & \lg & \lgg & LARES & OPTIS\\
\hline
$a$ (km) & 12270 & 12163 & 12270 & 29300\\
$e$ & 0.0045 & 0.014 & 0.04 & 0.478\\
$i$ (deg) & 110 & 52.65 & 70 & 63.4\\
$n$ (s$^{-1}$) & $4.643\times 10^{-4}$ & $4.710\times 10^{-4}$ &
$4.643\times 10^{-4}$ & $1.258\times 10^{-4}$\\
\br
\end{tabular}
\end{table}

The main characteristics of such a mission are the already
mentioned drag--free technique for OPTIS and the Satellite Laser
Ranging (SLR) technique for tracking. Today it is possible to
track satellites to an accuracy as low as a few mm. This may be
further improved in the next years.
\subsection{A modified OPTIS scenario}
It seems that an orbital configuration of OPTIS identical to that
of LARES of Table 1 would not be in dramatic contrast with the
requirements for the other originally planned tests of Special and
General Relativity. For example, the perigee height of LARES would
amount to 5400 km while the apogee height would be 6382 km, with
respect to Earth's surface. The difference in the gravitational
potential $\rp{\Delta U}{c^2}$ would be of the order of $3\times
10^{-11}$, which is only one order of magnitude smaller than the
one that could be obtained with the originally proposed OPTIS
configuration.
In the case of a satellite with spherically symmetric shape and a
small ratio cross sectional area to mass, such as the LAGEOS
satellites, the orbital perturbations due to the non-gravitational
perturbations are small and can be modelled with high accuracy.
Indeed, the orbits of these laser ranged satellites can be
modelled with root mean square of the orbital residuals, i.e. the
difference between the observed and the calculated orbital
elements, as little as about 1 cm over periods of about two weeks.
However, in the case of a satellite of complex shape such as OPTIS
we must rely on the drag-free system to reduce the effect of the
non-gravitational perturbations. For a satellite with relatively
small orbital eccentricity, the non-gravitational perturbations
are more effective on the perigee rate than on the nodal rate.
Indeed, over one orbital period, the total torque on the orbit due
to an acceleration $\bm A$, constant in magnitude and direction is
proportional to the eccentricity and the corresponding nodal
precession is proportional to $\frac{Ae}{na}$. A perturbing
acceleration $\bm A$, constant in magnitude and direction (in the
along-track direction), for example the nearly constant
along-track particle drag or a similar disturbing acceleration
$\bm A$, would produce a perigee precession proportional to
$\frac{A}{na}$. However a time--varying periodical perturbation
with period equal to the orbital period would produce a nodal
precession proportional to $\frac{A}{na}$ and a perigee precession
proportional to $\frac{A}{nae}$. Let us then estimate the order of
magnitude of the perigee and node perturbations of OPTIS--in the
LARES orbital configuration--due to the residual accelerations not
eliminated by the drag-free system. For OPTIS, in the frequency
range around $10^{-4}$ Hz and $10^{-3}$ Hz, corresponding to the
orbital period of OPTIS/LARES, the drag free system will reduce
the spurious accelerations down to $10^{-12}$ cm s$^{-2}$. Let us
then calculate the perigee and nodal rates induced by a periodical
acceleration of magnitude $10^{-12}$ cm s$^{-2}$ with period about
equal to the OPTIS orbital period. By integrating the Gauss
equation for perigee and node over one orbital period, we find
that a perturbing acceleration of $10^{-12}$ cm s$^{-2}$ with
frequency $\frac{2 \pi}{P}$, where $P$ is the orbital period of
OPTIS, would produce a perigee precession of the order of 0.2 mas
yr$^{-1}$ and a nodal precession of about 0.02 mas yr$^{-1}$.
Perturbing accelerations with frequencies $\frac{2 \pi}{P^*}$,
where the period $P^*$ is near the orbital period $P$, would also
produce a perigee precession with a comparable order of magnitude.
Since also sub-harmonics and higher harmonics would generate
perigee and nodal precessions and since there are various
non-gravitational perturbations with components near the orbital
frequency, i.e. direct solar and Earth albedo radiation pressure,
solar Yarkovsky, Yarkovsky-Rubincam effect, etc., we may assume
that the total effect of all the non-gravitational perturbations
with amplitude of the order of $10^{-12}$ cm s$^{-2}$, or less,
would at most induce a perturbation of a few percent of the
Lense-Thirring effect on the OPTIS perigee and of a fraction of
percent of the Lense-Thirring effect on the OPTIS node.

\subsection{A scenario without the perigee of
LAGEOS II} In (Iorio $et\ al$ 2002; Iorio 2003a) a multisatellite
combination of the orbital residuals of the nodes of LAGEOS,
LAGEOS II and LARES and the perigees of LAGEOS II and LARES has
been proposed in order to improve the obtainable accuracy of the
measurement of the Lense--Thiring effect. Such kind of
combinations are motivated by the need of reducing the impact of
the systematic error induced by the classical secular precessions
on the node and the perigee due to the mismodelling in the even
zonal harmonics of the geopotential (Iorio 2003b). Indeed, using
the combination of (Iorio $et\ al$ 2002) it would be possible to
cancel out the contributions of the first four even zonal
harmonics of the geopotential.

However, a possible weak point of this strategy is that it forces
to include in the combination the data of the perigee of LAGEOS
II, although with a weighing coefficient of the order of
$10^{-3}$. It is well known that the perigee of LAGEOS II is
severly affected by many non--gravitational perturbations
(Lucchesi 2002; 2002). Some of them, like the asymmetric
reflectivity, the solar Yarkovsky-Schach and the terrestrial
Yarkovsky-Rubincam effects, are of thermal origin and depend on
the temporal evolution of the LAGEOS II spin axis (Lucchesi 2002).
The key point is that, when (and if) OPTIS and/or LARES will be
finally launched, the spin may be chaotic and unpredictable. For
this purpose measurements of the LAGEOS II spin axis are and will
be performed using different techniques.
Moreover, it may happen that the averaging period necessary to
overcome the large variations in the perigee of LAGEOS II will
exceed the lifetime of the drag--free satellite which is related
to the amount of fuel and, consequently, to the mission budget.

In view of these considerations it would be meaningful to explore
the possibility of adopting a combination of orbital residuals
which does not include the perigee of LAGEOS II, even if it would
yield a systematic error due to the geopotential slightly less
favorable than the previous results. Indeed, when new, more
accurate Earth gravity models from the CHAMP (Pavlis 2000) and,
especially, GRACE missions (Ries $et\ al$ 2002) will be available,
the impact of the geopotential in the total error budget should be
dramatically reduced and should fall below negligible levels.

By assuming for OPTIS the same orbital configuration of LARES the
following combination yields high accuracy \eqi
\delta\dot\Omega^{\rm LAGEOS}+c_1\delta\dot\Omega^{\rm LAGEOS\
II}+c_2\delta\dot\Omega^{\rm OPTIS}+c_3\delta\dot\omega^{\rm
OPTIS} \ =61.3\mlt,\lb{combinopg}\eqf with \eqi c_1  \sim 3\times
10^{-3},\ c_2  \sim  9.9\times 10^{-1},\ c_3  \sim  1\times
10^{-3}\lb{ccF3}. \eqf In \rfr{combinopg} the quantity $\mlt$,
which is 0 in Galileo--Newton mechanics and 1 in General
Relativity\footnote{As explained in (Ciufolini 1996), $\mu_{\rm
LT}$ is also affected by the remaining, non--cancelled even zonal
harmonics of higher degree and by the small residuals of the
inclination.}, is the solved--for least square parameter which
accounts for the Lense--Thirring effect. The orbital residuals
$\delta\dot\Omega$ and $\delta\dot\omega$ would entirely absorb
the Lense--Thirring effect because the gravitomagnetic force would
be purposely set equal to zero in the force models of the orbital
processors, contrary to all the other classical and relativistic
accelerations which, instead, would be included in them. The
resulting gravitomagnetic signal would be a linear trend with a
slope of 61.3 mas yr$^{-1}$.

Note that, even if \rfr{combinopg} only cancels out the first
three even zonal harmonics, the systematic error due to the
remaining harmonics of higher degree amounts to
\eqi\left(\rp{\delta\mu_{\rm LT}}{\mu_{\rm LT}}\right)_{\rm even\
zonals}=4\times 10^{-4}\lb{erznopg}.\eqf The full covariance
matrix of EGM96 (Lemoine $et\ al$ 1998) up to degree $l=20$ has
been used. It can be shown that this result is also insensitive to
orbital injection errors in the inclination of OPTIS. Indeed, for
$i_{\rm OPT}$ ranging from 69 deg to 71 deg the corresponding
error varies from 0.04$\%$ to 0.06$\%$. It becomes
0.2$\%$--0.3$\%$ according to just the variance matrix of EGM96
used up to degree $l=20$ in a Root--Sum--Square (RSS) fashion. A
very pessimistic upper bound can be obtained by simply summing up
the absolute values of the individual errors induced by the
various even zonal harmonics. For EGM96 it amounts to
0.4$\%$--0.6$\%$. It may be interesting to get some insights about
the possible improvements which could be reached with the new
Earth gravity models from CHAMP and GRACE by using the data from
the recently released preliminary GGMC01C model which combines the
TEG-4 gravity model (Tapley $et\ al$ 2000) with the first data
from GRACE. It can be retrieved on the WEB at
http://www.csr.utexas.edu/grace/gravity/. According to a RSS
calculation with the variance matrix, the systematic relative
error due to the remaining harmonics of higher degree amounts
to\footnote{Note that, contrary to EGM96 in which the recovered
even zonal harmonics are highly correlated, the covariance
matrices of the GGM01C/S models are almost diagonal; then, in this
case, a RSS calculation should give a reliable estimate of the
error due to the mismodelling in the even zonal harmonics of
geopotential. Moreover, the released sigmas of the $J_l$
coefficients in the GGM01 models, although they are preliminary,
are not the mere formal, statistical errors but are tentatively
calibrated.} $3\times 10^{-5}$, with a pessimistic upper bound of
$6\times 10^{-5}$ obtained by summing up the absolute values of
the individual errors. Moreover, also the secular variations of
the even zonal harmonic coefficients of geopotential do not affect
the proposed combination. Indeed, it turns out that they can be
accounted for by an effective time rate (Eanes and Bettadpur 1996)
\eqi\dot J_2^{\rm eff}\sim\dot J_2+0.371\dot J_4+0.079\dot
J_6+0.006\dot J_8-0.003\dot J_{10}...\eqf whose magnitude is of
the order of $(-2.6\pm 0.3)\times 10^{-11}$ yr$^{-1}$, and
\rfr{combinopg} is designed in order to cancel out the effects of
just the first three even zonal harmonic coefficient of
geopotential. Finally, \rfr{combinopg} is affected neither by the
problem of the systematic bias of the even zonal harmonics due to
the Lense--Thirring signature (Ciufolini 1996). It consists of the
fact that that in the solutions of the various Earth gravity
models  General Relativity-and the Lense-Thirring effect itself-
is assumed to be true, so that the recovered $J_l$ are biased by
this a priori assumption. However, it turns out that, at least for
the LAGEOS satellites, such feature is mainly concentrated in the
first two--three even zonal harmonics.

With regard to the non-gravitational perturbations on the LAGEOS
satellites, only the contribution of the nodes of LAGEOS and
LAGEOS II, weighted by the small coefficients of \rfr{ccF3}, have
to be considered. This is quite relevant in the final error budget
because the nodes of the LAGEOS satellites, contrary to the
perigees of these laser-ranged satellites, are orbital elements
much less sensitive to the action of the non--gravitational
perturbations. With regard to the effect on the non--gravitational
perturbations on the OPTIS satellite we have already estimated the
impact of the residual accelerations, thus also according to the
evaluations of Table 2 and Table 3 of (Iorio $et\ al$ 2002), over
$T_{\rm obs}$ = 7 years \eqi\left(\rp{\delta\mu_{\rm LT}}{\mu_{\rm
LT}}\right)_{\rm NGP}\sim 3\times 10^{-3}.\lb{ngnpg}\eqf It should
be noted that the estimate of \rfr{ngnpg} is probably pessimistic.
Indeed, the periods of many time--dependent perturbations of the
nodes of LAGEOS and LAGEOS II, contrary to the perigee of LAGEOS
II, are far shorter than 7 years\footnote{For example, the period
of the tesseral ($m=1$) tidal perturbation $K_1$, which is one of
the most powerful perturbations affecting the node of LAGEOS,
amounts to 2.85 years (Iorio 2001). Of course, the semisecular
orbital tidal perturbations induced by the 9--year and 18.6--year
tides do not affect \rfr{combinopg} because they are $l=2$, $m=0$
perturbtions and, consequently, are cancelled out.}, so that would
be possible to adopt a $T_{\rm obs}$ of just a few years during
which it should be possible to save fuel and fit and remove the
small time--varying non-gravitational signals affecting the nodes
of the LAGEOS satellites or average them out. Moreover, the
perigee of OPTIS would have an impact of the order of $10^{-4}$
for a residual, unbalanced acceleration $\delta A=10^{-12}$ cm
s$^{-2}$. Last but not least, the impact of the perigee of LAGEOS
II, difficult to be modeled at a high level of accuracy, is
absent.
So, the total final systematic error budget in measuring the
Lense--Thirring effect with \rfr{combinopg} should be of the order
of $1\%$.

When more robust and confident Earth gravity solutions will be
available in the near future, the need for canceling out as many
even zonal harmonics as possible will be less stringent than now
and, then, it could be possible to discard both the perigee of
LAGEOS II and of OPTIS as well. So, a three--nodes combination
could be considered. Indeed, by using the nodes of LAGEOS, LAGEOS
II (with a coefficient of $3\times 10^{-3}$) and OPTIS (with a
coefficient of 9.9$\times 10^{-1}$) the relative error due to the
static part of geopotential, according to the variance matrix of
GGM01C (RSS calculation) would be $3\times 10^{-5}$, with an upper
bound of $6\times 10^{-5}$. The slope of the gravitomagnetic
signal would be 61.4 mas yr$^{-1}$. In this case, since the nodes
are insensitive to the larger Post--Newtonian gravitoelectric
force which, instead, affects the perigee, the result of such test
would be independent of the inclusion of it into the force
models\footnote{However, it should be pointed out that the
gravitoelectric pericentre advance has already been tested in the
gravitational field of Sun with interplanetary ranging at a
$10^{-2}-10^{-3}$ precision level (Will 1993) ; this level of
accuracy in its knowledge would have a negligible impact in a
measurement of the Lense--Thirring effect with an observable like
that of \rfr{combinopg} with the coefficient $c_3$ given by
\rfr{ccF3}. }. With the three--nodes combination it should not be
too optimistic to predict a total systematic error of the order
of, or less than 1$\%$ over a time span of a few years.
\subsection{The impact of the observational errors}
Of crucial importance for the presented scenario would be, of
course, the quality of the OPTIS tracking and orbital data
reduction which should be, if possible, of the same level as that
of the LAGEOS satellites. At present, the technique for the OPTIS
orbital reconstruction has not yet been established, according to
(L${\rm \ddot{a}}$mmerzahl $et\ al$ 2001). If it will be finally
decided to adopt the SLR (Satellite Laser Ranging) approach, a too
high altitude of OPTIS might reduce the quality of the recovered
orbit due to calibration problems and to the variable number of
photons reflected back to the ranging stations. However, within
the suggested project ASTROD (Huang $et\ al$ 2002) research is
under way with the goal to make precise phase coupling to very
weak laser beams. Therefore, the accuracy of laser ranging might
improve in the near future also for larger distances. Moreover,
NASA has successfully tested laser ranging from Earth to Mars some
years ago. The concept is based on a slightly modified system from
what SLR community now uses on LAGEOS and the other existing
geodetic satellites, but it could be applied to OPTIS without any
problem. The fact that the satellite is rotating should not
present a problem, as long as there is some ahead of launch
planning to deal with this. Nearly all satellites spin, and not
all of them are spherical such as LAGEOS, but they are still
tracked. It will simply be considered when the CCR arrays to
account for that spinning will be designed.

Another point to consider is that the large eccentricity of the
originally proposed OPTIS configuration, contrary to the other
existing geodetic satellites of LAGEOS--type, would not allow for
a uniform coverage of the laser--ranged data in the sense that
certain portions of the orbit would remain poorly tracked. 

In conclusion, the scenario of \rfr{combinopg}, with OPTIS instead
of LARES in its orbital configuration, would yield a very accurate
measurement of the Lense--Thirring effect at a level of the order
of, or better than 1$\%$. Instead, the peculiar originally
proposed orbital configuration of OPTIS may pose some problems for
the orbital reconstruction with the currently available SLR
technique.
%

%
\section{The problem of the eccentricity}
Perhaps the major point of conflict between the original designs
of the OPTIS and LARES missions is represented by the eccentricity
$e$ of the orbit of the spacecraft. Indeed, while for the
gravitational redshift test, given by \rfr{reds}, a relatively
large value of $e$  is highly desirable, the originally proposed
LARES orbit has a smaller eccentricity. The point is twofold: on
one hand, it is easier and cheaper, in terms of requirements on
the performances of the rocket launcher, to insert a satellite in
a nearly circular orbit, and, on the other, the present status of
the ground segment of SLR would assure a uniform tracking of good
quality for such kind of orbits.

However, the originally proposed OPTIS mission implies the use of
a rocket ARIANE 5 to insert the spacecraft into a GTO orbit and,
then, the use of a kick motor. Moreover, it may be reasonable to
assume that, when OPTIS/LARES will be launched, the network of SLR
ground stations will have reached a status which will allow to
overcome the problem of reconstructing rather eccentric orbits to
a good level of accuracy.

Then, a reasonable compromise between the OPTIS and LARES
requirements could be an eccentricity of, say, $e=0.1$. In that
case \rfr{reds} yields a gravitational redshift of $\frac{\Delta U
}{c^2}=7.3\times 10^{-11}$ about 2 - 3 times smaller than in the
original OPTIS proposal. Accordingly, the accuracy of the tests
concerning the gravitational red shift will be worse be a factor 2
to 3. With regard to the Lense--Thirring effect, it turns out
that, for the combination without the perigee of LAGEOS II of
\rfr{combinopg}, the error due to the even zonal harmonics of
geopotential would amount to 1.5$\%$, according to the diagonal
part only of the covariance matrix of EGM96 up to degree $l=20$
(RSS calculation). The sum of the absolute values of the
individual terms yields an upper bound of the order of 3$\%$.
However, the forthcoming Earth gravity models from CHAMP and GRACE
will greatly improve also such estimates. Indeed, the variance
matrix of the very preliminary GGM01C model yields an error of
0.02$\%$ (RSS calculation) and a pessimistic upper bound of the
order of 0.04$\%$ from the sum of the absolute values of the
individual errors.

It may also be interesting to note that the originally proposed
observable of the LAGEOS--LARES mission, i.e. the sum of their
nodes, would be affected by such change in the eccentricity of the
orbit of LARES at a 5$\%$--7$\%$ level, according to diagonal part
only of the covariance matrix of EGM96 up to degree $l=20$ (RSS
calculation), with an upper bound of the order of 12$\%$--16$\%$
from the sum of the absolute values of the individual errors. A
RSS calculation with the variance matrix of GGM01C yields a
0.7$\%$--1.5$\%$ level of percent error and an upper bound of
1$\%$--2.2$\%$ from the sum of the absolute values of the
individual errors.

With a larger eccentricity the impact of the non--gravitational
perturbations would be reduced and, on the other hand, the
accuracy of the measurement of the Lense-Thirring effect on the
OPTIS/LARES perigee would be increased. For example, the amplitude
of the non--gravitational perturbations would reduce to 0.1 mas
yr$^{-1}$ and, for $\delta r^{\rm exp}\sim 1$ cm over a certain
time span, the observational error in the perigee would amount to
just 1 mas.
\section{Would an OPTIS/LARES mission still be useful in measuring the Lense--Thirring effect? }
In view of the expected improvements of the even zonal harmonics
of geopotential by the new forthcoming Earth gravity models from
CHAMP and GRACE, which should ameliorate our knowledge of,
especially, the mid--high degree part of the even zonal harmonics
spectrum, it seems legitimate to ask if the corresponding
improvements in the obtainable accuracy of measurements of the
Lense--Thirring effect with the currently existing laser--tracked
satellites would make unnecessary a dedicated mission to this goal
like LARES.

In (Iorio and Morea 2004) preliminary estimates with the recently
released EIGEN-2 CHAMP--only and GGM01C GRACE--based models have
been carried out. It turned out that the systematic error due to
the mismodelling in the even zonal harmonics of geoptential of the
combination proposed in (Ciufolini 1996), which involves the nodes
of LAGEOS and LAGEOS II and the perigee of LAGEOS II, would be
$\leq 3\%$, while the error of the combination put forth in (Iorio
and Morea 2004), based on the nodes of LAGEOS and LAGEOS II only,
would be $\leq 18\%$, according to the variance matrix of GGM01C.
The use of the nodes of other existing SLR satellites (Ajisai,
Starlette, Stella) would induce a systematic error $\leq 123\%$.
Note that such figures represent a conservative, pessimistic upper
bound obtained by adding the sum of the absolute values of the
individual errors.

If further improvements in our knowledge of the terrestrial
gravitational field will come from new, more robust and reliable
solutions from GRACE, as it is expected, it seems reasonable to
suppose that the systematic errors due to geopotential will reduce
to a some percent level for the nodes--only combination of (Iorio
and Morea 2004) and below the 1$\%$ level for the three-elements
combination involving also the perigee of LAGEOS II. Much will
depend on the magnitude of the improvements of the low--degree
even zonal harmonics $J_4,\ J_6, J_8,...$ which will be obtained.
The LAGEOS satellites are not particularly sensitive to the even
zonal harmonics of degree higher than $l=12$, so that an increased
accuracy in knowing them would be of relatively little usefulness
for the measurement of the Lense--Thirring effect with LAGEOS and
LAGEOS II.

In view of this considerations, and by noting that,of course, also
the proposed measurement with OPTIS/LARES would benefit from the
improved knowledge of the terrestrial gravitational field, it is
possible to conclude that the level of accuracy in measuring the
gravitomagnetic force obtainable by implementing the OPTIS mission
will remain far higher than that could be reached by simply
reanalyzing the data of the existing SLR satellites. Needless to
say that the latter approach would not allow to perform all the
other tests of Special Relativity and Post--Newtonian gravitation
which, in turn, could be implemented by the originally proposed
LARES with a lower accuracy or could not be implemented at all.

\section{Conclusions}
In this paper we have shown that it would be possible to perform a
very accurate measurement of the Lense-Thirring effect with the
orbital data of the proposed OPTIS drag--free satellite, in
addition to the other previously proposed tests of Special
Relativity and Post--Newtonian gravitation. OPTIS is currently
under serious examination by the German Aerospace Agency. In order
to use the orbital data of OPTIS for precise tests of relativistic
gravity it would be necessary, first of all, to carry onboard some
SLR passive retroreflectors in order to reconstruct with great
accuracy its orbit. To this aim, it turns out that the originally
proposed orbital configuration of OPTIS, based on a highly
eccentric orbit with a perigee of 10000 km, would not probably be
well suited for, e.g., adequate SLR tracking. It would be better
to adopt a LAGEOS--like orbit; it turns out that the orbital
configuration of, e.g., the proposed LARES would not be in
dramatic contrast with the requirements of the other relativistic
experiments originally planned for OPTIS. With such a choice it
would be possible to adopt a linear combination of the orbital
residuals of the nodes of LAGEOS, LAGEOS II and OPTIS in order to
measure the Lense--Thirring effect with a total systematic error
that should be of the order of 1$\%$ or, perhaps, better.  Such
orbital test of Post--Newtonian gravitomagnetism require
observational temporal intervals of some years in order to average
out or fit and remove various time--dependent perturbations of
gravitational and, especially, non--gravitational origin acting on
the Keplerian orbital elements to be adopted in the analysis. So,
it is of the utmost importance that the lifetime of the drag--free
apparatus of OPTIS, which would not be a passive, spherical,
geodetic satellite of LAGEOS--type, would not be shorter than the
time span of the data analysis. However, on one hand the
technology to be adopted should meet such requirements yielding
lifetimes of the order of 10 years, on the other, the exclusion of
the perigee of LAGEOS II, which is affected by some gravitational
and non--gravitational perturbations with long periods, assures
that not too long observational time spans would be needed.
Finally, a comparison between the obtainable accuracy in measuring
the Lense--Thirring effect with OPTIS and the one that could be
obtained by simply reanalyzing the data of the existing SLR
satellites with the new Earth gravity models shows that the former
approach would yield unrivalled results.

In conclusion, the use of OPTIS for measuring the Lense--Thirring
effect is feasible: in regard to this goal, the best choice would,
probably, be to adopt the orbital configuration of LARES. It would
not too seriously affect the obtainable accuracy in the
gravitational red-shift test which is particularly sensitive to
the orbital eccentricity. However, a compromise solution could
also be adopted. The same observables as LARES could be employed
with better results thanks to the active drag--free apparatus to
be employed on OPTIS; indeed, LARES would be a totally passive
satellite. Moreover, in addition to the Lense--Thirring effect,
OPTIS would allow to perform many other tests of Special
Relativity and Post--Newtonian gravitation.
\ack L Iorio is grateful to L Guerriero for his support while at
Bari. E C Pavlis gratefully acknowledges NASA's support through
the Co-operative Agreement NCC 5-339. Special thanks also to M C W
Sandford for his important suggestions and information about the
drag--free technology.
\References
\item [] Ciufolini I and Wheeler J A 1995 {\it Gravitation and Inertia}
(New York: Princeton University Press)
\item [] Ciufolini I 1986 Measurement of
\leti\ drag on high-altitude, laser ranged artificial satellites
{\it Phys. Rev. Lett.} {\bf 56} 278--81
\item [] Ciufolini I 1996 On a new method to measure the
gravitomagnetic field using two orbiting satellites {\it Il Nuovo
Cimento A} {\bf 109} 1709--20
\item [] Ciufolini I 1998 The Concept of the LARES experiment in {\it
LARES Phase--A Study} (Universit\`{a} La Sapienza: Rome) pp 16--33
\item [] Ciufolini I 2002 Test of general relativity: 1995--2002
measurement of frame--dragging {\it Proceedings of the Physics in
Collision conference}, Stanford, California, June 20--22, 2002,
{\it Preprint} gr-qc/0209109
\item [] Ciufolini I, Pavlis E C, Chieppa F, Fernandes-Vieira E and
P{\'{e}}rez-Mercader J 1998 Test of General Relativity and
Measurement of the Lense-Thirring Effect with Two Earth Satellites
{\it Science} {\bf 279} 2100--3
\item [] Dittus H, L\"{a}mmerzahl C, Peters A and Schiller S 2002
OPTIS--a satellite test of Special and General Relativity, paper
H0.1-1-0017-02 presented at {34th COSPAR Scientific Assmbly},
Houston, TX, 10th October-19th October, 2002
\item [] Eanes R J and Bettadpur S V 1996
Temporal variability of Earth's gravitational field from satellite
laser ranging in {\it Global Gravity Field and its Temporal
Variations (IAG Symp. Ser. 116)} ed Rapp R H, Cazenave A and Nerem
R S (New York: Springer) pp 30-41
\item [] Everitt C W F and other members of the Gravity Probe B team
2001 Gravity Probe B: Countdown to Launch in \textit{Gyros,
Clocks, Interferometers...:Testing Relativistic Gravity in Space}
ed L\"{a}mmerzahl C, C W F Everitt and F W Hehl (Springer Verlag:
Berlin) pp. 52--82
\item [] Huang T, L${\rm\ddot{a} }$mmerzahl C, Ni W--T, R${\rm
\ddot{u}}$diger A and Zhang Y--Z (guest eds.) 2002 {\it
Proceedings of the First International ASTROD Symposium on Laser
Astrodynamics, Space Test of Relativity and Gravitational-Wave
Astronomy } in {\it Int. J. Mod. Phys. D} {\bf 11} 947--1158
\item [] Iorio L 2001 Earth tides and Lense-Thirring effect
\textit{Celest. Mech. \& Dyn. Astron.} {\bf 79} 201--30
\item [] Iorio L 2003a A reassessment of the systematic gravitational
error in the LARES mission {\it Gen. Rel. Grav.} {\bf 35} 1263--72
\item [] Iorio L 2003b The impact of the static part of the Earth's
gravity field on some tests of General Relativity with Satellite
Laser Ranging {\it Celest. Mech. \& Dyn. Astron.} {\bf 86} 277--94
\item []
Iorio L and Morea A 2004 The impact of the new Earth gravity
models on the measurement of the Lense--Thirring effect {\it Gen.
Rel. Grav.} in press {\it Preprint gr-qc/0304011}
\item [] Iorio L, Lucchesi D M and Ciufolini I 2002 The LARES mission
revisited: an alternative scenario {\it Class. Quantum Grav.} {\bf
19} 4311--25
\item [] L$\ddot{\rm a}$mmerzahl C, Dittus H, Peters A and Schiller S
2001 OPTIS--a satellite--based test of Special and General
Relativity {\it Class. Quantum Grav.} {\bf 18} 2499--508
\item [] L{\"a}mmerzahl C and Dittus H 2002 Fundamental physics in
space: A guide to present projects {\it Ann. Physik} {\bf 11}
95--150
\item [] L\"ammerzahl C, Ahlers G, Ashby N, Barmatz M, Biermann P L,
Dittus H , Dohm V, Duncan R, Gibble K, Lipa J, Lockerbie N,
Mulders N and Salomon S 2004 Experiments in Fundamental Physics
scheduled and in development for the ISS, to appear in {\it Gen.\
Rel.\ Grav.}
\item [] Lemoine F G, Kenyon  S C, Factor J K, Trimmer R G,
Pavlis N K, Chinn D S, Cox C M, Klosko S M, Luthcke S B, Torrence
M H, Wang Y M, Williamson R G, Pavlis E C, Rapp R H and Olson T R
1998 The Development of the Joint NASA GSFC and the National
Imagery Mapping Agency (NIMA) Geopotential Model EGM96
NASA/TP-1998-206861
\item [] Lense J and Thirring H 1918 \"{U}ber den Einfluss der
Eigenrotation der Zentralk{\"{o}}rper auf die Bewegung der
Planeten und Monde nach der Einsteinschen Gravitationstheorie {\it
Phys. Z.} {\bf 19} 156--63, translated by Mashhoon B, Hehl F W and
Theiss D S 1984 On the Gravitational Effects of Rotating Masses:
The Thirring-Lense Papers
 {\it Gen. Rel. Grav.} {\bf 16} 711--50
\item [] Lockerbie N, Mester J C, Torii R, Vitale S and Worden P W
2001 {STEP}: A Status Report in {\it Gyros, Clocks, and
Interferometers: Testing Relativistic Gravity in Space} ed
L{\"a}mmerzahl C, Everitt C W F and Hehl F W (Springer--Verlag:
Berlin) pp. 213--47
\item [] Lucchesi D 2001 Reassessment of the error modelling of
non-gravitational perturbations on LAGEOS II and their impact in
the Lense-Thirring determination. Part I {\it Plan. and Space
Sci.} {\bf 49} 447--63
\item [] Lucchesi D 2002 Reassessment of the error modelling of
non-gravitational perturbations on LAGEOS II and their impact in
the Lense--Thirring determination. Part II {\it Plan. and Space
Sci.} {\bf 50} 1067--100
\item [] Maleki L and Prestage J 2001 Space{T}ime Mission: Clock Test
of Relativity at Four Solar Radii in {\it Gyros, Clocks, and
Interferometers: Testing Relativistic Gravity in Space} ed
L{\"a}mmerzahl C, Everitt C W F and Hehl F W (Springer--Verlag:
Berlin) pp. 369--80
\item [] Mashhoon B, Iorio L and Lichtenegger H I M 2001 On the
gravitomagnetic clock effect {\it Phys. Lett. A} {\bf 292} 49--57
\item [] Nobili A M, Bramanti D, Polacco E, Roxburgh I W, Comandi G
and Catastini G 2000 `Galileo Galilei' (GG) small-satellite
project: an alternative to the torsion balance for testing the
equivalence principle on Earth and in space {\it Class. Quantum
Grav.} {\bf 17} 2347--9
\item [] Pavlis E C 2000 Geodetic Contributions to Gravitational
Experiments in Space in {\it Recent Developments in General
Relativity} ed Cianci R, Collina R, Francaviglia M and Fr${\rm
\acute{e}}$  P  (Springer--Verlag: Milan) pp. 217--33
\item [] Pound R V and Rebka G A 1960 Apparent weight of Photons {\it
Phys.\ Rev.\ Lett.} {\bf 4} 337--41
\item [] Ries J C, Eanes R J, Tapley B D and Peterson G E 2002
Prospects for an Improved Lense-Thirring Test with SLR and the
GRACE Gravity Mission {\it Proceedings of the 13th International
Laser Ranging Workshop}, Washington DC, October 7-11, 2002 {\it
Preprint
http://cddisa.gsfc.nasa.gov/lw13/lw$\_${proceedings}.html$\#$science}
\item [] Ries J C, Eanes R J and  Tapley B D 2003 Lense-Thirring
Precession Determination from Laser Ranging to Artificial
Satellites in {\it Nonlinear Gravitodynamics. The Lense--Thirring
Effect} ed Ruffini R and Sigismondi C  (World Scientific:
Singapore) pp. 201--11
\item [] Sanders A J, Alexeev A D, Allison S W, Antonov V, Bronnikov K
A, Campbell J W, Cates M R, Corcovilos T A, Earl D D, Gadfort T,
Gillies G T, Harris M J, Kolosnitsyn N I, Konstantinov M Yu,
Melnikov V N, Newby R J, Schunk R G and Smalley L L 2000 Project
{SEE} ({S}atellite {E}nergy {E}xchange): an international effort
to develop a space--based mission for precise measurements of
gravitation {\it Class. Quantum Grav.} {\bf 17} 2331--46
\item [] Sch\"afer G 2003 Gravitomagnetic effects, preprint,
University of Jena, to appear in {\it Proceedings of 1st HYPER
symposium} held in Paris, Nov. 2002
\item [] Schiff L I 1960 On Experimental Tests of the General Theory
of Relativity {\em Am. J. Phys.} {\bf 28} 340--3
\item [] Tapley B D,  Chambers d D P, Cheng M K, Kim M C, Poole S and
Ries J C 2000 The TEG-4 Earth Gravity Field Model, paper presented
at 25th EGS General Assembly, Nice, France, April 2000
\item [] Touboul P 2001a {MICROSCOPE}, testing the equivalence
principle in space {\it Comptes Rendus de l'Acad. Sci. S\'erie IV:
Physique Astrophysique} {\bf 2} 1271--86
\item [] Touboul P 2001b Space Accelerometer: Present Status in {\it
Gyros, Clocks, and Interferometers: Testing Relativistic Gravity
in Space} ed L{\"a}mmerzahl C, Everitt C W F and Hehl F W
(Springer--Verlag: Berlin.) pp. 273--91
\item [] Turyshev S, Shao M and Nordvedt K 2003a
The Laser Astrometric Test of Relativity (LATOR) Mission invited
talk presented at {\it 2003 NASA/JPL Workshop on Fundamental
Physics in Space}, Oxnard, CA, April 14-16, 2003, {\it Preprint}
gr-qc/0311020
\item [] Turyshev S, Williams J G, Nordvedt K, Shao M and Murphy T
W 2003b 35 Years of Testing Relativistic Gravity: Where do we go
from here? invited talk presented at {\it 302.WE-Heraeus-Seminar:
"Astrophysics, Clocks and Fundamental Constants"}, Bad Honnef,
Germany, June 16-18, 2003, {\it Preprint} gr-qc/0311039
\item [] Vessot R F C, Levine M W, Mattison E M, Blomberg E L,
Hoffmann T E, Nystrom G U, Farrel B F, Decher R, Eby P B, Baughter
C R, Watts J W, Teuber D L and Wills F D 1980 Test of Relativistic
Gravitation with a Space--Borne Hydrogen Maser {\it Phys.\ Rev.\
Lett.} {\bf 45} 2081--4
\item [] Will C M 1993 \textit{Theory and Experiment in Gravitational
Physics} 2nd edition (Cambridge University Press: Cambridge)
\endrefs
\end{document}